
\input harvmac


\long\def\optional#1{}
%
%

\def\dag{\dagger}     
\fontdimen16\tensy=2.7pt\fontdimen17\tensy=2.7pt 



%
%

\def\CM{{\cal M}}

%
%
\def\ts{\thinspace}
\def\ra{\rightarrow}

\def\ra{\rightarrow}
\def\ol{\overline}

\def\Few{F_\pi}

\def\W+-{W^\pm}
\def\Z0{Z^0}

\def\qbq{\ol{Q}Q}

\def\Ms+-{M^2_\pm}

\def\M+-{M_\pm}

\def\tro{\rho_T}

\def\tpi{\pi_T}
\def\tpiqq{\pi_{\ol Q Q}}
\def\tpiuu{\pi_{\ol U U}}
\def\tpidd{\pi_{\ol D D}}

\def\tpilq{\pi_{\ol L Q}}
\def\tpinu{\pi_{\ol N U}}
\def\tpieu{\pi_{\ol E U}}
\def\tpind{\pi_{\ol N D}}
\def\tpied{\pi_{\ol E D}}
\def\tpiql{\pi_{\ol Q L}}

\def\kslash{\raise.15ex\hbox{/}\kern-.57em k}
\def\pslash{\raise.15ex\hbox{/}\kern-.57em p}
%
%
%

\def\gev{\rm {GeV}}

\def\pb{{\rm pb}}

\def\ecm{\sqrt{s}}
\def\rshat{\sqrt{\shat}}
\def\shat{\hat s}


\def\simge{\mathrel{%
   \rlap{\raise 0.511ex \hbox{$>$}}{\lower 0.511ex \hbox{$\sim$}}}}
\def\simle{\mathrel{
   \rlap{\raise 0.511ex \hbox{$<$}}{\lower 0.511ex \hbox{$\sim$}}}}

\def\slashchar#1{\setbox0=\hbox{$#1$}           
   \dimen0=\wd0                                 
   \setbox1=\hbox{/} \dimen1=\wd1               
   \ifdim\dimen0>\dimen1                        
      \rlap{\hbox to \dimen0{\hfil/\hfil}}      
      #1                                        
   \else                                        
      \rlap{\hbox to \dimen1{\hfil$#1$\hfil}}   
      /                                         
   \fi}                                         %
\def\emiss{\slashchar{E}}
\def\etmiss{\emiss_T}
%
%

%
%

%
%
%
%


\newwrite\ffile
\def\footend{ \def\foot{\global\advance\ftno by1\chardef\wfile=\ffile
$^{\the\ftno}$\ifnum\ftno=1\immediate\openout\ffile=foots.tmp\fi%
\immediate\write\ffile{\noexpand\smallskip%
\noexpand\item{f\the\ftno:\ }\pctsign}\findarg}
\def\listrefs{\vfill\eject\immediate\closeout\ffile\parindent=20pt
\centerline{{\bf Footnotes}}\bigskip\input foots.tmp
\vfill\eject\immediate\closeout\rfile\parindent=20pt
\baselineskip14pt\centerline{{\bf References}}\bigskip\frenchspacing%
\input refs.tmp\vfill\eject\nonfrenchspacing}
} 

\global\newcount\figno \global\figno=1
\newwrite\ffile
\def\pfig#1#2{Fig.~\the\figno\nfig#1{#2}}
\def\nfig#1#2{\xdef#1{Fig. \the\figno}%
\ifnum\figno=1\immediate\openout\ffile=figs.tmp\fi%
\immediate\write\ffile{\noexpand\item{\noexpand#1\ }#2}%
\global\advance\figno by1}

%
%
\def\tfig#1{Fig.~\the\figno\xdef#1{Fig. \the\figno}\global\advance\figno by1}

%
%
%
\def~{\ifmmode\phantom{0}\else\penalty10000\ \fi}
\def\lae{\raise-.5ex\vbox{\hbox{$\; <\;$}\vskip-2.9ex\hbox{$\; \sim\;$}}}
\def\gae{\raise-.5ex\vbox{\hbox{$\; >\;$}\vskip-2.9ex\hbox{$\; \sim\;$}}}


\def\fun#1#2{\lower3.6pt\vbox{\baselineskip0pt\lineskip.9pt
  \ialign{$\mathsurround=0pt#1\hfil##\hfil$\crcr#2\crcr\sim\crcr}}}

%
%


\overfullrule=0pt
\def\half{\textstyle{ { 1\over { 2 } }}}
\def\third{\textstyle{ { 1\over { 3 } }}}

\def\Mh{M_{\eta_T}}
\def\tbt{\ol t t}
\def\bbb{\ol b b}

\def\qbq{\ol q q}
\def\pbp{\ol p p}
\def\QbQ{\ol Q Q}

\def\Mtt{\CM_{\ol t t}}
\def\jet{{\rm jet}}
\def\jets{{\rm jets}}
\def\pt{p_T}
\def\et{E_T}

\def\Dzero{{\rm D}\slashchar{0}}
\def\myfoot#1#2{{\baselineskip=14.4pt plus 0.3pt\footnote{#1}{#2}}}
\overfullrule=0pt

\Title{\vbox{\baselineskip12pt\hbox{FERMILAB--PUB--94/007--T}
\hbox{BUHEP--94--1}}}
{Multiscale Technicolor and Top Production}

\centerline{Estia Eichten\myfoot{$^{\dag }$}{eichten@fnal.gov}}
\smallskip\centerline{Fermi National Accelerator Laboratory}
\centerline{P~.O.~Box 500 Batavia, IL 60510}
\smallskip
\centerline{and}
\smallskip
\centerline{Kenneth Lane\myfoot{$^{\ddag }$}{lane@buphyc.bu.edu}}
\smallskip\centerline{Department of Physics, Boston University}
\centerline{590 Commonwealth Avenue, Boston, MA 02215}
\vskip .3in

\centerline{\bf Abstract}

Pair--production of heavy top quarks at the Tevatron Collider is
significantly enhanced by the color--octet technipion, $\eta_T$, occurring
in multiscale models of walking technicolor. We discuss $\tbt$ rates for
$m_t = 170\,\gev$ and $\Mh = 400-500\,\gev$. Multiscale models also have
color--octet technirho states in the mass range $200-600\,\gev$ that appear
as resonances in dijet production and technipion pair--production.

\bigskip

\Date{1/94}

\vfil\eject

Over the past year there has been intense searches for the top quark by
the CDF
\ref\cdfref{CDF Collaboration:
F.~Abe, et al., FERMILAB--PUB--Conf--93/212--E~(1993);
T.~Chikamatsu, ``Search for the Top Quark in the
Dilepton Channel at CDF", and
M.~Contreras, ``Top Search in the Lepton plus Jets
Channel at CDF", in
``Proceedings of the $9^{\rm th}$ Topical Workshop
in $\pbp$ Collider Physics", Tsukuba~(1993), ed.~by K.~Kondo.}
and $\Dzero$
\ref\d0ref{D0 Collaboration: M.~Strovink, ``Proceedings of the International
Europhysics Conference on High Energy Physics", Marseille~(1993), eds. J.~Carr
and M.~Perrottet;
M.~Fatyga, ``Search for the Top Quark at $\Dzero $
(in the di-lepton channels)" and
H.~Greenlee, ``Search for the Top Quark in the
Single Lepton Plus Jets Channel at CDF",  in
``Proceedings of the $9^{\rm th}$ Topical Workshop
in $\pbp$ Collider Physics", Tsukuba~(1993), ed.~by K.~Kondo.}
collaborations using data obtained during the recent  high--luminosity run of
the Tevatron Collider at Fermilab.\foot{The integrated luminosity collected by
CDF is $22\,\pb^{-1}$; for
$\Dzero$ it is $15\,\pb^{-1}$.} Two main signatures have been sought:
(1)~Events with two isolated high--energy leptons ($e^\pm$ and $\mu^\pm$)
and large missing transverse energy ($\etmiss$); (2)~Events with an
isolated lepton associated with multijets ($\ge 3$) and large $\etmiss$.
Both are signatures of the standard--model processes expected at the
Tevatron: QCD production of $\tbt$ with each top--quark decaying as $t \ra
W b \ra (\ell + \etmiss + \jet)$ or 3~jets. So far, a clear signal for this
standard top--quark production has not emerged. However, CDF has reported
observation of several events of the second type which also have one
of the jets identified as arising from a $b$--quark.
The jets in these events have very large $\et$.
These events have {\it no}
appreciable standard--model physics source other than $\tbt$
production~\ref\gieleref{J.~M.~Benlloch, K.~Sumorok, and W.~Giele,
``Possibilities of
Discovering a Heavy Top Quark in the Lepton-Multijet Channel",
FERMILAB--Pub--93/276--T~(1993) and references therein.}.
In this Letter, we presume that these $\tbt$ candidate events are, in
fact, real. We assume that the top--quark mass is 170~GeV, near the central
value extracted from precision electroweak measurements at LEP
\ref\lepref{W.~Hollik, ``Status of the Electroweak Standard
Model",  presented at the XVI~International Symposium on Lepton-Photon
Interactions, Cornell University, Aug.~10-15, 1993, Ithaca, NY.}.

The purpose of this Letter is to point out that the $\tbt$ rates and
associated pair--mass and momentum distributions measured in these Tevatron
experiments may probe flavor physics which is beyond the standard model.
Top--quark production can be significantly modified from QCD expectations by
the resonant production of {\it colored}, flavor--sensitive scalar
particles with mass in the range $400-500\,\gev$.\foot{C.~Hill and S.~Parke
recently considered the effect on the $\tbt$ rate of color--singlet
and octet vector resonances that couple strongly to top
quarks\ref\hp{C.~Hill and S.~Parke, FERMILAB--Pub--93/397--T.}.} In
particular, we emphasize that the color--octet technipion, $\eta_T$,
expected to occur in multiscale models
\ref\multia{K.~Lane and E.~Eichten, Phys.~Lett.~{\bf 222B} (1989)~274.},
\ref\multib{K.~Lane and M~V.~Ramana, Phys.~Rev.~{\bf D44} (1991)~2678.}
of walking technicolor
\ref\tcrefs{S.~Weinberg, Phys.~Rev.~{\bf D13}(1976)~974;
{\it ibid}, {\bf D19} (1979)~1277\semi
L.~Susskind, Phys.~Rev.~{\bf D20} (1979)~2619.},
\ref\etcrefs{S.~Dimopoulous and L.~Susskind, Nucl.~Phys.~{\bf B155}
(1979)~237\semi
E.~Eichten and K.~Lane, Phys.~Lett.~{\bf 90B} (1980)~125.},
\ref\wtc{B.~Holdom, Phys.~Rev.~{\bf D24} (1981)~1441;
 Phys.~Lett.~{\bf 150B} (1985)~301 \semi
T.~Appelquist, D.~Karabali and L.~C.~R. Wijewardhana,
Phys.~Rev.~Lett.~{\bf 57} (1986) 957~;
T.~Appelquist and L.~C.~R.~Wijewardhana, Phys.~Rev.~{\bf D36} (1987)~568
\semi  K.~Yamawaki, M.~Bando and K.~Matumoto, Phys.~Rev.~Lett.~{\bf 56},
(1986)~1335 \semi
T.~Akiba and T.~Yanagida, Phys.~Lett.~{\bf 169B} (1986)~432.}
can easily double the $\tbt$ rate. The $\eta_T$ occurs in technicolor
models which have color--triplet techniquarks
\ref\etatrefers{E.~Farhi and L.~Susskind Phys.~Rev.~{\bf D20}
(1979)~3404\semi
S.~Dimopoulos, Nucl.~Phys.~{\bf B168} (1980)~69 \semi
T.~Appelquist and J.~Terning, Yale and Boston University Preprint
YCTP-P21-93, BUHEP-93-23 (1993).}.
The production in hadron collisions via gluon fusion
of a ``standard'' $\eta_T$---the one
occurring in a one--family technicolor model and having decay constant $F =
123\,\gev$ and nominal couplings to quarks and gluons---has been extensively
discussed elsewhere
\ref\ehlq{E.~Eichten, I.~Hinchliffe, K.~Lane and C.~Quigg,
Rev.~Mod.~Phys.~{\bf 56} (1984)~579; {\it ibid}, Phys.~Rev.~{\bf 34}
(1986)~1547.}
\ref\etatprod{T.~Appelquist and G.~Triantaphyllou, Phys.~Rev.~Lett.~{\bf
69} (1992)~2750.}.
We shall see that the standard $\eta_T$ with $\Mh \sim 400\,\gev$ increases
the $\tbt$ rate by only 15\%. Because of uncertainties in QCD corrections
to the standard--model $\tbt$ rate, this is unlikely to be observable. In
multiscale models, however, the $\eta_T$ decay constant is much smaller, $F
\sim 20-40\,\gev$. For $\Mh = 400-500\,\gev$, this small decay constant is
what accounts for a measurably larger $\tbt$ rate.

If an $\eta_T$ with multiscale dynamics produces an excess of $\tbt$
events, then there also must appear color--octet technirhos, $\tro$, which
have flavor--blind couplings to quarks and gluons. The models discussed in
Refs.~\multia, \multib\ indicate that they have mass in the range
$200-600\,\gev$. It is quite possible that at least one of these $\tro$
decays predominantly into $gg$ and $\qbq$, and appears as a resonance in
ordinary dijet production. In addition to the $\eta_T$, there will be other
flavor--sensitive scalars---technipions, $\tpi$---which are color octets
and, possibly, color--triplets (leptoquarks). They have masses in the same
general range as the $\eta_T$ and the $\tro$.\foot{We expect that these
technipions are so heavy that the decay $t \ra \pi_T b$ is forbidden.} They
are strongly pair--produced in the Tevatron Collider experiments and their
rates are enhanced if the decays $\tro \ra \tpi \tpi$ are allowed. Thus,
the hallmark of the new physics signalled by excess $\tbt$ events is the
appearance of colored technihadrons: scalars that are flavor--sensitive and
vectors that may be flavor--blind. In addition, there will be
color--singlet technihadrons, some decaying into $W$ and $Z$--bosons. These
were discussed in Ref.~\multia. We urge searches for all these states as
soon as possible.

In standard ETC models, the mass of the $\eta_T$ arises mainly from QCD
interactions (see S.~Dimopoulos in Ref.~\etatrefers). For example, suppose
that the technicolor group is $SU(N_{TC})$, that the technifermions
transform according to the fundamental representation, $\bf{N_{TC}}$, and
that they consist of one doublet of QCD--color triplet techniquarks, $Q =
(U,D)$, and $N_D-3$ doublets of color--singlet technileptons, $L_i = (N_i,
E_i)$. Then, the mass of the $\eta_T$ has been estimated to be $\Mh = 240
\ts \sqrt{N_D/N_{TC}}\,\gev$. In walking technicolor models, proposed to
suppress flavor--changing neutral currents while maintaining reasonable
quark masses, there is a large and probably dominant ETC contribution to
$\Mh$~\multib.

The $\eta_T$ is expected to decay predominantly into $\tbt$, $\bbb$ and
$gg$. So long as the $\eta_T$ is an approximate Goldstone boson, the
amplitude for $\eta_T \ra gg$ is reliably calculated from the
Adler--Bell--Jackiw triangle anomaly. For one doublet of techniquarks in
the ${\bf N_{TC}}$ representation of $SU(N_{TC})$,\foot{See, e.g.,
Ref.~\ehlq\ and references therein. This amplitude may be modified by a
form factor for the process $\eta_T \ra g \rho_T$; $\rho_T \ra g$. We do
not expect this effect to change our conclusions significantly.}
\eqn\Aetagg{A(\eta^a_T(p) \ra g_b(p_1) \ts g_c(p_2)) =
{\alpha_s(\Mh) \ts N_{TC} \ts  d_{abc} \over {2 \pi \sqrt{2}\ts  F_Q}} \ts
\epsilon_{\mu\nu\lambda\rho} \epsilon_1^\mu \epsilon_2^\nu p_1^\lambda
p_2^\rho \ts.}
Here, $F_Q$ is the decay constant of technipions in the $\QbQ$ sector. If the
only technifermions are techniquarks and technileptons comprising $N_D$
doublets, then $F_Q \cong \Few/\sqrt{N_D}$ where $\Few = 246\,\gev$.
The amplitude for $\eta_T \ra \qbq$ is more dependent on the details of
the particular ETC model. The coupling to $\qbq$
is expected to be approximately $m_q/F_Q$.
To take into account ETC-model dependence, we introduce a dimensionless
factor $C_q$, expected to be not much different from one, and write
\eqn\Aetaqq{A(\eta^a_T(p) \ra q(p_1) \ts \ol q(p_2)) =
{C_q \ts m_q \over{F_Q}} \ts
\ol u_q(p_1) \ts \gamma_5 \ts {\lambda_a \over{2}} \ts v_q (p_2) \ts.}
Then the $\eta_T$ decay rates are
\eqn\etarates{\eqalign{
\Gamma(\eta_T \ra gg) &= {5 \alpha_s^2 \ts N_{TC}^2 \ts \Mh^3  \over {384 \ts
\pi^3 \ts F_Q^2}} \ts; \cr
\Gamma(\eta_T \ra \qbq) &= {C_q^2 \ts m_q^2 \ts \Mh \ts \beta_q \over {16 \pi
F_Q^2}} \ts.\cr}}
Here, $\beta_q = \sqrt{1 - 4 m_q^2/\Mh^2}$. For a one-family ETC model with
$\Mh = 400\,\gev$, the $\eta_T$ decay rates are $\Gamma(\eta_T \ra \tbt) =
8.0\,\gev$, $\Gamma(\eta_T \ra \bbb) = 0.013\,\gev$, and $\Gamma(\eta_T \ra
gg) = 0.28\,\gev$.\foot{The parameters used here are $m_t = 170\,\gev$,
$m_b = 5\,\gev$, $\alpha_s(\Mh) = 0.1$, $N_{TC} = 4$, $F_Q = 123\,\gev$,
and $C_b = C_t = 1$.}

At the Tevatron Collider with $\sqrt{s} = 1800\,\gev$, and with
large $m_t$, standard $\tbt$ production is dominated by light $\qbq$
annihilation
\ref\qcdref{P.~Nason, S.~Dawson, and R.~K.~Ellis, Nucl.~Phys. {\bf B303}
(1988)~607; W.~Beenakker, H.~Kuijf, W.~L.~van~Neerven and
J.~Smith, Phys.~Rev. {\bf D40} (1989)~54.}.
Using the EHLQ Set~1 distribution functions to compute the $\tbt$ rate
from the lowest--order cross sections, we find $\sigma(\pbp \ra \tbt) =
3.6\,\pb$ for $m_t = 170\,\gev$. Next--to--leading--log corrections
and soft--gluon resummation
\ref\resum{E.~Laenen, J.~Smith and W.~L.~Van Neerven, Nucl.~Phys.~{\bf B369}
(1992)~543; {\it ibid}, FERMILAB--Pub--93/270--T.}
give rates which are 50\% larger than these in this general top--mass range.
Accordingly, throughout this paper we scale our computed $\tbt$
cross sections by a factor of~1.5.
So long as the $\eta_T$ is relatively narrow, the process $gg \ra \eta_T \ra
\tbt$ does not interfere (in lowest order) with the purely QCD production
processes. The differential
cross section at subprocess center--of--mass energy $\rshat$
is given by\foot{In Eq.~(5), we are using partially $\shat$-dependent widths,
with $\beta_t = \sqrt{1 - 4m_t^2/\shat}$ and $\alpha_s = \alpha_s(\rshat)$.}
\eqn\etatsig{
{d \hat \sigma(gg \ra \eta_T \ra \tbt) \over {dz}} =
{\pi \over{4}} \ts {\Gamma(\eta_T \ra gg)\ts \Gamma(\eta_T \ra \tbt) \over
{(\shat - \Mh^2)^2 + \shat \ts \Gamma^2(\eta_T) }} \ts.}
Here, $z = \cos \theta$, where $\theta$ is the subprocess c.~m.~scattering
angle. Combining this formula with the lowest--order QCD cross sections,
and using the parameters assumed above, we find a total $\tbt$ rate of
4.1~pb, to which $\eta_T$ contributes only 0.54~pb. We assume that
higher-order QCD corrections increase $\hat \sigma(gg \ra \eta_T \ra \tbt)$
by the same amount as they do the QCD cross sections.\foot{For standard
$\tbt$ production, higher order QCD corrections to the $gg \ra \tbt$
amplitude are significantly larger than to the $\qbq \ra \tbt$
amplitude~\resum. Since the production of the $\eta_T$ is in the symmetric
color--octet $gg$ channel, our assumption may be conservative.} Then, the
standard $\eta_T$ probably has no observable effect on $\tbt$ production.

To understand why multiscale technicolor
implies a much larger $\eta_T \ra \tbt$ rate, let us examine $\sigma(\pbp \ra
\eta_T \ra \tbt)$. For a relatively narrow $\eta_T$, it is given by
\eqn\etatrate{
\sigma(\pbp \ra \eta_T \ra \tbt) \simeq {\pi^2 \over {2 s}} \ts
{\Gamma(\eta_T \ra gg) \ts \Gamma(\eta_T \ra \tbt) \over{\Mh \ts
\Gamma(\eta_T)}} \ts \int_{-Y_B}^{Y_B} d y_B \ts z_0 \ts
f_g^{(p)} (\sqrt{\tau} e^{y_B}) \ts f_g^{(p)}(\sqrt{\tau} e^{-y_B}) \ts.}
In Eq.~\etatrate, $f_g^{(p)}$ is the gluon distribution function in the
proton, $\tau = \Mh^2 /s$, $y_B$ is the boost rapidity of the subprocess
frame, and $z_0$ is the maximum value of $z = \cos \theta$ allowed by
kinematics and fiducial cuts \ehlq. The key point of Eq.~\etatrate\ is
that, unless the $\eta_T \tbt$ strength factor
$C_t \simle 0.2$, the cross section is simply proportional to
$\Gamma(\eta_T \ra gg)$ and the form of this decay rate is fairly
model-independent: it depends only on the technicolor and color
representations of the $\eta_T$ and on $F_Q$. In our case, it is
proportional to $N_{TC}^2/F_Q^2$. Thus, the small decay constant of the
$\eta_T$ in multiscale technicolor implies a large $\sigma(\pbp \ra \eta_T
\ra \tbt)$.

The multiscale model studied in Ref.~\multib\ has many theoretical
and phenomenological difficulties
(not the least of which is obtaining a large top--quark mass
unless one invokes near-critical extended technicolor interactions
\ref\setca{T.~Appelquist, T.~Takeuchi, M.~B.~Einhorn, L.~C.~R.~Wijewardhana,
 Phys.~Lett.~{\bf 220B} (1989)~223; T.~Takeuchi, Phys.~Rev.~{\bf D40}
(1989)~2697
\semi V.~A.~Miransky and K.~Yamawaki, Mod.~Phys.~Lett.~{\bf A4}
(1989)~129}).
It is not our intention here to advocate adoption of the model in
detail. However, to focus our discussion, we extract from it
that there is one doublet of techniquarks, perhaps one or more doublets of
technileptons, and the associated spectrum of technipions and technirhos at
a scale that is relatively low compared to the electroweak breaking scale.
Details of the high--scale technifermions, those most directly responsible
for electroweak symmetry breaking, are not important for our
considerations.

In the remainder of this Letter, we generally assume that $F_Q = 40\,\gev$.
We consider two cases for the $\eta_T \ts \tbt$ coupling: $C_t = 1$ and
$C_t = \third$, both with $m_t = 170\,\gev$. The number of technicolors
will be $N_{TC} = 4$ and we use $\Mh = 400-500\,\gev$ to study
the effect of the $\eta_T$ mass on the distributions of the
$\tbt$ invariant mass, $\Mtt$

Figures~1 and 2 show the invariant mass distributions, $d\sigma(\pbp \ra
\tbt)/d \Mtt$, to lowest order in QCD for $\Mh = 400\,\gev$ and $C_t =
1$~and~$\third$. The total cross section as well as its QCD and $\eta_T$
components are shown. No cut is put on the top-quark rapidity. The $\eta_T$
widths and integrated cross sections are summarized in Table~1.\foot{It is
clear from the table that, for the parameters we used, the narrow--width
approximation of Eq.~\etatrate\ is only approximately satisfied.} The decay
constant we chose, $F_Q = 40\,\gev$, is at the upper end of the values
found in the multiscale model calculations. Thus, an $\eta_T$ in this mass
range easily can double the $\tbt$ production rate. In the absence of the
$\eta_T$, we calculate the mean $\Mtt$ for a 170~GeV top quark to be
430~GeV. The closer the $\eta_T$ is to the $\tbt$ threshold, the lower is
this $\langle \Mtt \rangle$.

These invariant mass distributions and rates convey a qualitative
impression of the effect of varying the $\eta_T$ mass and width. Because
the main production mechanisms at the Tevatron energy, $\qbq \ra \tbt$ and
$gg \ra \eta_T \ra \tbt$, are central, the $\pt$ distributions for the top
quarks are expected to have a shape similar to Figs.~1--2, with $\pt(t) =
\vert  \sum_{t \ra \jets}\vec\pt(\jet)\vert \simeq 0.5 m_t$. Detailed event
and detector simulations are needed to determine the best variables to test
for the presence of the $\eta_T$ in the existing data and in
higher--luminosity samples.

If the $\eta_T$ of multiscale technicolor exists, there will also be
color--octet $\tro$ and $\tpi$ in the same general mass region and they
will, in principle, be observable in the Tevatron experiments. Their
signatures are more dependent on the details of the model than the $\eta_T$
signatures are. We briefly discuss two general cases, distinguished by
whether techni--isospin ($I_T$) breaking is negligible or not. In both
cases we assume that there is at least one doublet of technileptons $L =
(N,E)$, so that there are color--triplet (leptoquark), as well as octet,
technipions. We denote the two types by $\tpiql$, $\tpilq$ and $\tpiqq$,
respectively.

If $I_T$--breaking is small, the techniquark hard masses satisfy $m_Q \equiv
m_U \cong m_D$. Similarly, $m_L \equiv m_N \cong m_E$. Then, all $\tpiqq$ are
degenerate, as are all leptoquarks and all octet $\tro$. If we ignore
QCD contributions, their masses are given by~\multia,\multib\
\eqn\masses{\eqalign{
M^2_{\pi_{\ol Q Q}} &\simeq 2 m_Q \ts \langle \overline Q Q
\rangle_{\Lambda_Q} /F_Q^2 \ts, \cr
M^2_{\pi_{\ol Q L}} &\simeq (m_Q + m_L) \ts \langle
\overline Q Q \rangle_{\Lambda_Q} /F_Q^2 \ts, \cr
M_{\rho_T} &\simeq 2\ts (m_Q + \Lambda_Q) \ts. \cr}}
Here, $\Lambda_Q$ is the techniquark condensation scale; we relate it to
the $\eta_T$ decay constant by $\Lambda_Q \simeq F_Q \ts (\half
M_\rho/f_\pi) = 165\,\gev$ for $F_Q = 40\,\gev$. The techniquark condensate
(renormalized at $\Lambda_Q$) is estimated to be $\langle \overline Q Q
\rangle_{\Lambda_Q} \simeq 4 \pi F^3_Q$. These mass formulae are true
regardless of the size of $I_T$--breaking. They imply simple sum rules
which can be employed should candidates for the $\tpi$ and $\tro$ ever be
found. For example, note that $M_{\pi_{\ol Q L}} \ge M_{\pi_{\ol Q
Q}}/\sqrt{2}$. For $\Mh = 400\,\gev$, we obtain $m_Q \simeq 160\,\gev$ and
$M_{\rho_T} \simeq 650\,\gev$. The color--octet technipion decay channels
of $\tro$ are closed. The leptoquark channels are also closed {\it if} $m_L
> 0.32 \ts m_Q \simeq 50\,\gev$.

If the $\tro$ lies below the two--technipion threshold, it decays mainly
into $\ol q q$ and $gg$ dijets. With $\tro$--coupling parameters chosen as
in Ref.~\multib, the $\tro$ is narrow, $\Gamma(\tro \ra 2\ts \jets) \simeq
12\,\gev$.\foot{In Ref.~\multib\ we used $\Gamma(\rho^a_T \ra g^a \ra
gg):\Gamma(\rho^a_T \ra g^a \ra \ol q_i q_i) = 3:1$. We expect that if
observable ETC modifications of these results occur, they will be
flavor--symmetric. We thank R.~S.~Chivukula for bringing this issue to our
attention.} We calculated the excess dijet cross section in the vicinity of
$\CM_{\rm jj} = 650\,\gev$ to be $2.5-1.0\,\pb$. This sits on a background
of $1.0\,\pb$. Radiative corrections have not been applied. The range of
variation in the signal includes an estimate of
the effect of jet--energy resolution, which is about 5\%
for CDF at $\CM_{\rm jj} = 650\,\gev$
\ref\cdfjets{F.~Abe, et al. Phys.~Rev.~{\bf D48} (1993)~999.}.
Observation of this dijet resonance will require very high integrated
luminosity at the Tevatron.

The $\tro$ width will be dominated by the leptoquark decay channels if they
are open. The leptoquarks are themselves expected to decay as $\tpinu \ra
\ol \nu t$, $\tpieu \ra \tau^+ t$, $\tpind \ra \ol \nu b$, and $\tpied \ra
\tau^+ b$. Again, the cross sections are only in the few~pb range,
depending on the number of technileptons and the masses of the leptoquarks.

Consider now the case that $I_T$--breaking is appreciable. The $\tro$ and
$\tpi$ will be approximately ideally--mixed states. For example, the
electrically--neutral color--octets appear as $\ol U U$ and $\ol D D$
states instead of $(\ol U U + \ol D D)/\sqrt{2}$ and $(\ol U U - \ol D
D)/\sqrt{2}$.  Thus, there are now two ``$\eta_T$'' produced in $gg$
fusion: $\tpiuu$ decaying mainly to $\tbt$ and $\tpidd$ decaying mainly to
$gg$ (unless the factor $C_b \gg 1$). We expect $m_U > m_D$, hence
$M_{\pi_{\ol U U}} > M_{\pi_{\ol D D}}$. The effect on the $\eta_T$ decay
amplitudes is to multiply $A(\eta_T \ra gg)$ by $1/\sqrt{2}$ and $A(\eta_T
\ra \qbq) $ by $\sqrt{2}$, changes that can be hidden in the magnitude of
$F_Q$ and $C_q$.  There will be no measurable enhancement of the dijet rate
due to $\pbp \ra \tpidd \ra gg$.

In Ref.~\multib, it was found that the $\rho_{\ol U U}$ generally was above
$\tpi \tpi$ threshold. Whether the lighter $\rho_{\ol D D}$ lay above or
below the threshold was dependent on calculational details. To illustrate
one possible scenario, we have considered the case $M_{\rho_{\ol D D}}
\simeq 375\,\gev <$~$2M_{\pi_T}$ and $M_{\rho_{\ol U U}} \simeq 500\,\gev
>$~$2 M_{\pi_T}$.\foot{The technipion masses were taken to be $M_{\ol U U}
= 400\,\gev$, $M_{\ol U D} = 325$, $M_{\ol D D} = 225$,  $M_{\pi_{\ol N U}}
= 300$, $M_{\pi_{\ol N D}} = M_{\pi_{\ol E U}} = 250$, and $M_{\pi_{\ol E
D}} = 200$.} The signal and background dijet cross sections are shown in
Fig.~3 and, with a dijet mass resolution of about 7\%, in Fig.~4. Also
shown in Fig.~3 are the $\tro$ signal in the $\bbb$ channel. The jet
rapidities were required to be less than~0.7. Such a tight cut is necessary
to observe the central--region signal.

The $\rho_{\ol D D}$ true width is about 3~GeV. The integral, from 360~GeV
to 400~GeV, over the resonant cross section is 70~pb, while the background
is 50~pb (that is, a signal--to--background ratio of $S/B =
20\,\pb/50\,\pb$). The CDF jet--energy resolution deteriorates this $S/B$
significantly. The integrals, from 325~GeV to 425~GeV, over the total and
background cross sections in Fig.~4 are 145~pb and 130~pb, respectively.
The $S/B$ in the (unsmeared) $\bbb$~signal is much higher than for the
total dijet cross section: $2.7\,\pb/0.5\,\pb$.\foot{We thank Frank Paige
for suggesting that the $\bbb$ channel would have a better $S/B$ than the
gluon and light quark channels.} However, to take account of this
enhancement with an integrated luminosity of $50-100\,\pb^{-1}$ requires a
$b$-jet identification and reconstruction efficiency of at least 25\%.
Finally, in this case the $\rho_{\ol U U}$ resonance is practically
invisible in the dijet signal. It must be sought in $\rho_{\ol U U} \ra
\tpi \tpi$. Typical rates are discussed in Ref.~\multib. Efficient
heavy-flavor ($t$, $b$, $\tau$) tagging will be essential.

In this Letter, we have re\"emphasized that multiscale technicolor has
low--energy degrees of freedom that can significantly enhance the rates of
heavy--flavor processes under study at the Tevatron Collider. The
color--octet $\eta_T$ of multiscale technicolor, with its small decay
constant, $F_Q = 30-40\,\gev$, can easily double the top--quark production
rate and skew its distributions. Color--octet $\tro$ may lie below
technipion threshold and appear as narrow resonances in dijet production.
If $\tro \ra \tpi\tpi$ occurs, the technipions may be sought via their
expected decay to heavy quarks and leptons. If the basic ideas of
multiscale technicolor underlie the physics of electroweak symmetry
breaking, a broad spectrum of measurements will be needed at the Tevatron
to limit scenarios and help pin down basic parameters. The discovery of the
high--scale technihadrons most directly linked to electroweak symmetry
breaking must await high--luminosity multi--TeV colliders. However, the
Tevatron Collider experiments may herald the true beginning of our
understanding of flavor physics.

We thank Sekhar Chivukula, Chris Quigg and Elizabeth Simmons for their
thoughtful comments and careful reading of the manuscript. We also
thank Walter Giele and Keith Ellis for discussions. EE's~research is
supported by the Fermi National Accelerator Laboratory, which is operated
by Universities Research Association, Inc., under
Contract~No.~DE--AC02--76CHO3000. KL'~research is supported in part by the
Department of Energy under Contract~No.~DE--FS02--91ER40676 and by the
Texas National Research Laboratory Commission under Grant~No.~RGFY93-278.
KL~also thanks the Fermilab Theory Group for its hospitality during the
final stage of this paper's preparation.

\listrefs

\centerline{\vbox{\offinterlineskip
\hrule\hrule
\halign{&\vrule#&
  \strut\quad#\hfil\quad\cr
height4pt&\omit&&\omit&&\omit&&\omit&&\omit&&\omit&&\omit&\cr
&\hfill $\Mh$ \hfill&&\hfill$C_t$ \hfill&&\hfill
 $\Gamma(\eta_T \ra \tbt)$\hfill&&\hfill
$\Gamma(\eta_T \ra gg)$  \hfill&&\hfill $\sigma_{\rm tot}(\tbt)$\hfill&&
$\hfill \sigma_{\eta_T}(\tbt)$\hfill&&
\hfill $\langle \Mtt \rangle$ \hfill &\cr
height4pt&\omit&&\omit&&\omit&&\omit&&\omit&&\omit&&\omit&\cr
\noalign{\hrule}
height4pt&\omit&&\omit&&\omit&&\omit&&\omit&&\omit&&\omit&\cr
&$400$&&\hfill$1$\hfill&&\hfill$76$\hfill
&&\hfill$2.88$\hfill&&
\hfill$11.4$\hfill&&\hfill $5.87$\hfill&&\hfill $410$\hfill&\cr
height4pt&\omit&&\omit&&\omit&&\omit&&\omit&&\omit&&\omit&\cr
&$400$&&\hfill$\third$\hfill&&\hfill$8.4$\hfill
&&\hfill$2.88$\hfill&&
\hfill$11.5$\hfill&&\hfill $5.96$\hfill&&\hfill $415$\hfill&\cr
height4pt&\omit&&\omit&&\omit&&\omit&&\omit&&\omit&&\omit&\cr
&$450$&&\hfill$1$\hfill&&\hfill$106$\hfill
&&\hfill$3.99$\hfill&&
\hfill$9.21$\hfill&&\hfill $3.70$\hfill&&\hfill $425$\hfill&\cr
height4pt&\omit&&\omit&&\omit&&\omit&&\omit&&\omit&&\omit&\cr
&$450$&&\hfill$\third$\hfill&&\hfill$11.8$\hfill
&&\hfill$3.99$\hfill&&
\hfill$8.38$\hfill&&\hfill $2.86$\hfill&&\hfill $435$\hfill&\cr
height4pt&\omit&&\omit&&\omit&&\omit&&\omit&&\omit&&\omit&\cr
&$500$&&\hfill$1$\hfill&&\hfill$132$\hfill
&&\hfill$5.35$\hfill&&
\hfill$7.98$\hfill&&\hfill $2.46$\hfill&&\hfill $430$\hfill&\cr
height4pt&\omit&&\omit&&\omit&&\omit&&\omit&&\omit&&\omit&\cr
&$500$&&\hfill$\third$\hfill&&\hfill$14.6$\hfill
&&\hfill$5.35$\hfill&&
\hfill$6.90$\hfill&&\hfill $1.39$\hfill&&\hfill $440$\hfill&\cr
height4pt&\omit&&\omit&&\omit&&\omit&&\omit&&\omit&&\omit&\cr}
\hrule}}
\medskip
\centerline{TABLE 1: $\eta_T$ widths, $\tbt$ cross sections in $\pbp$
collisions at 1800~GeV, and mean $\Mtt$.}
\medskip
{\noindent The top quark mass is 170~GeV. The $\eta_T$ decay constant is
$F_Q = 40\,\gev$. Masses and widths are in~GeV; cross sections are in
picobarns. QCD~radiative corrections have been estimated by multiplying
cross sections by~1.5.}
\vfil\eject

\centerline{\bf Figure Captions}
\bigskip

\item{[1]} The $\tbt$ invariant mass distribution for $\Mh = 400\,\gev$ and
$C_t = 1$ in $\pbp$ collisions at $\ecm = 1800\,\gev$. The QCD (dotted
curve), $\eta_T \ra \tbt$ (dashed), and total (solid) rates have been
multiplied by 1.5 as explained in the text.

\medskip

\item{[2]} The $\tbt$ invariant mass distribution for $\Mh = 400\,\gev$ and
$C_t = \third$ in $\pbp$ collisions at $\ecm = 1800\,\gev$. Curves are
labeled as in Fig.~1.

\medskip

\item{[3]} The invariant mass distributions for dijets (upper curves) and
$\bbb$ (lower curves) in $\ol p p$ collisions at $\ecm = 1800\,\gev$. The
solid curves include the $\rho_{\ol D D}$ and $\rho_{\ol U U}$ resonances
at 375 and 500~GeV. The dashed curves show the standard QCD distributions.
Radiative corrections have not been applied.

\medskip

\item{[4]} The dijet mass distributions as in the upper curves of Fig.~3,
except that a uniform resolution smearing of $\Delta \CM / \CM = 7\%$ has
been applied.

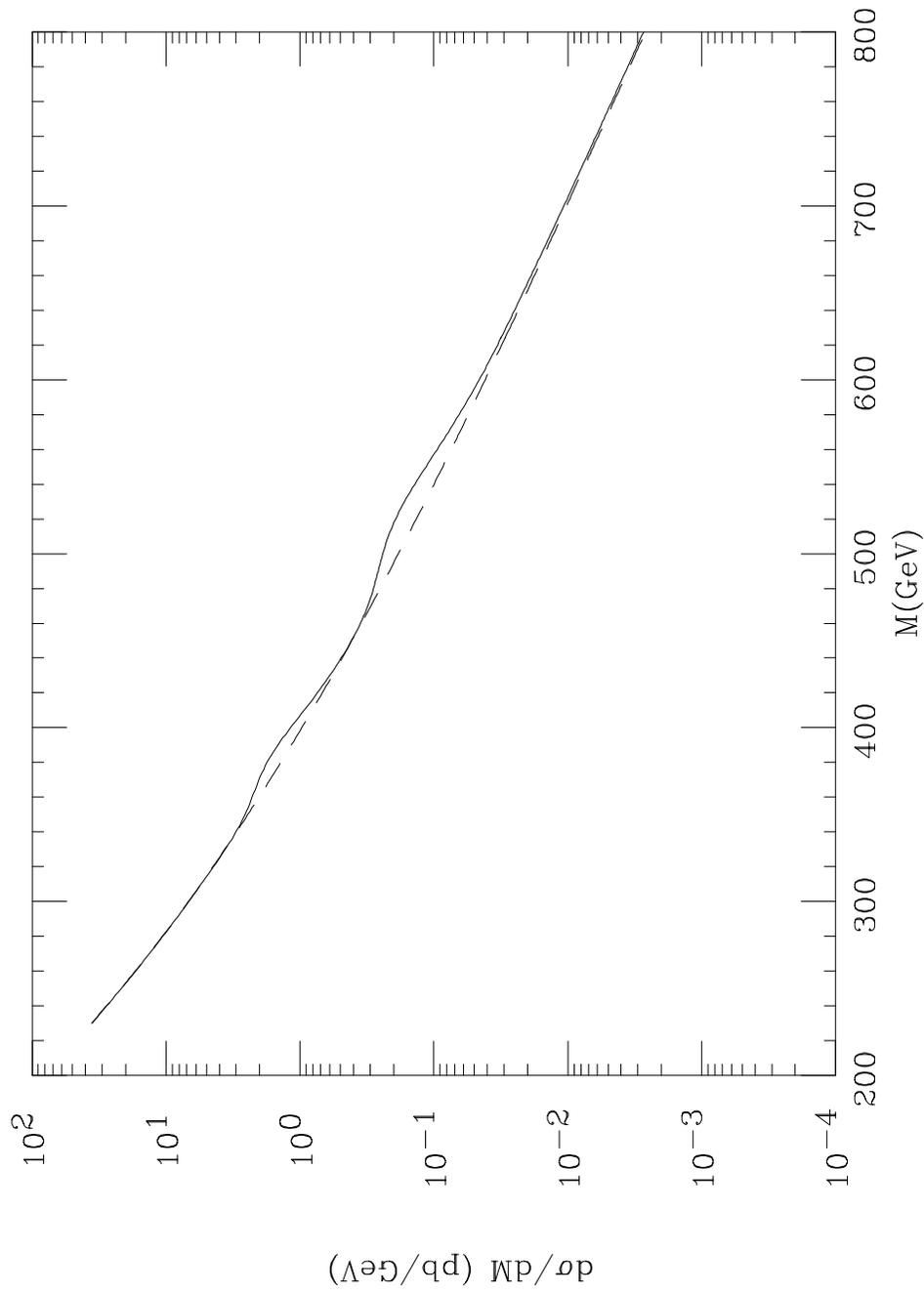
\bye